\documentclass[11pt]{article}
\usepackage[dvips,usenames]{color}
\usepackage[dvips]{pstcol} 

\usepackage{graphicx}

\def \be{\begin{displaymath}}
\def \ee{\end{displaymath}}              

\def \ben{ \begin{equation} }
\def \een{ \end{equation}   }            

\def \bea{\begin{eqnarray*}}             
\def \eea{\end{eqnarray*}}

\def \bean{\begin{eqnarray}}             
\def \eean{\end{eqnarray}}

\def \nn{\nonumber}

\def \Pr{\mbox{Pr}}

\def \span{ \mbox{span} }
\def \tr{ \mbox{{\rm tr}}}

\def \fie {\varphi}                    
\def \eps{\varepsilon}

\def \cl#1{ {\cal #1} }               
\def \b#1{ \mathbf{#1} }               

\def \Ref#1{(\ref{#1})}

\def \inv{ ^{-1} }
\def \dag{\;\!\!^\dagger}

\def \tenpow#1{^{\otimes #1}}                  

\def \invb#1 { \frac{1}{#1} }
\def \bra#1{\langle #1 |}                      
\def \ket#1{| #1 \rangle }                     
\def \pro#1{ \ket{ #1 }\!\bra{ #1 } }          
\def \ketbra#1#2{ \ket{ #1 }\!\bra{ #2 } }     

\def \sp#1{ \langle #1 \rangle }               

\def \cav#1{ [ #1 ] }                          
\def \scav#1{ \left[ #1 \right] }              

\def \fr#1#2{ \frac{#1}{#2} }

\def \norm#1{\parallel \! #1 \! \parallel}

\def \tn#1{\norm{#1}_{tr}}           

\def \fn#1{\norm{#1}_{F}}            

\def \scrbox#1{{\scriptsize\mbox{#1}}}

\begin{document}
\input epsf

\title{A random-coding based proof for the quantum coding theorem}
\author{Rochus Klesse\footnote{Email address: rk@thp.uni-koeln.de}\\[0.5cm]
\centerline{ {\sl Universit\"at zu K\"oln, Institut f\"ur Theoretische
    Physik, Z\"ulpicher Str. 77,}}\\
\centerline{ {\sl  D-50937 K\"oln, Germany}} }
\date{October 16, 2007}
\maketitle
\begin{abstract}
We present a proof for the quantum channel coding theorem
which relies on the fact that a randomly chosen code space typically
is highly suitable for quantum error correction. 
In this sense, the proof is close to Shannon's original treatment of
information transmission via a noisy classical channel.
\end{abstract}

\section{Preliminaries}\label{sec-preliminaries}

\subsection{Quantum channel}\label{sub-quantum-channel}

In the theory of information transmission the information is ascribed
to the configuration of a physical system, and the transmission is
ascribed to the dynamical evolution of that configuration under the
influence of an in general noisy environment. It is therefore customary
to characterize an information carrying system solely by its
configuration space, and to consider its intrinsic dynamics as part of
the transmission.

In a quantum setting we identify a system $Q$ with its
Hilbert space, denoted by the same symbol $Q$.  Its
dimension $|Q|$ will be always assumed to be finite.
The system's configuration is a quantum state described by a density
operator $\rho$ in $\cl B(Q)$, the set of bounded operators on $Q$.

The process of information transmission can be any dynamics of an open
quantum system $Q$ according to which an initial input state $\rho$
evolves to a final output state $\rho'$, defining in this way the
operation of a quantum channel $\cl N$
\footnote{For an introduction into the theory of quantum information
  see e.g.~\cite{NC00,Key02}. }.
Mathematically, $\cl N$ is a completely
positive mapping of $\cl B(Q)$ onto itself, or, when we admit that the
system may change to an other system $Q'$ during the course of transmission,
onto $\cl B(Q')$, the set of bounded operators on $Q'$, 
\bea
\cl N \: : \: \cl B(Q) & \rightarrow & \cl B(Q') \nn\\
\rho &\mapsto & \rho' = \cl N(\rho)\:. 
\eea

According to Stinespring's theorem \cite{Sti55} the operation of the
channel 
can be always understood as an isometric transformation followed by a
restriction \cite{Kra83,Key02}. That is,
one always finds an ancilla system $E$ with $|E| \ge 1$ 
and an isometric operation $V \: : \: Q \to Q'E$
such that for all states $\rho$
\be
\cl N (\rho) = \tr_{E} V \rho V\dag\:, 
\ee
where $\tr_{E}$ denotes the partial trace over $E$. In the following
we refer to this construction as Stinespring representation.
An elementary physical interpretation of it becomes obvious in the
case $Q=Q'$.  
Here one can find a unitary operator $U$ on $Q E$ and a state
vector  $ \ket{ \fie_{E} } \in E$ such that 
$
V \ket \psi = U \ket \psi \otimes \ket{ \fie_{E} }  
$
for all state vectors $\ket \psi \in Q$. Interpreting $U$ as time
evolution operator of the joint system $Q E$, an initial state 
$\rho \otimes \fie_E$, where $\fie_E = \pro{ \fie_E}$, will
evolve to final state $U\rho \otimes \fie_E U\dag$. Its partial trace with
respect to $E$ yields indeed $\cl N(\rho)$ as the reduced density operator
for $Q$, 
\be
\tr_E U\rho \otimes \fie_E U\dag = \tr_E V\rho V\dag = \cl N(\rho)\:.
\ee 

If we fix an orthonormal basis $\ket{1}, \dots \ket{N}$ of $E$
\footnote{Since we assumed the dimensions $|Q|$ and $|Q'|$ to be finite
also the ancilla $E$ can be chosen to be of finite dimension $|E|=N$.}, the 
Stinespring representation can be rewritten more explicitly in an operator sum
as
\be
\cl N(\rho) = \sum_{k=1}^N A_k \rho A_k\dag\:, 
\ee
where Kraus operators $A_1, \dots, A_N \: : \: Q \to Q'$ are 
defined by 
$
A\ket \psi  :=  \sp{k|V|\psi}
$ \cite{Kra83,NC00,Key02}.
Because $V$ is an isometry the Kraus operators satisfy the
completeness relation $\sum_{k=1}^N A_k\dag A_k = \b 1_{Q}$. 

Below, we will often have to refer to the number of Kraus operators 
of a channel $\cl N$ in a certain operator-sum representation, which,
of course, equals the dimension $|E|$ of the ancilla $E$ in the
corresponding Stinespring representation. It is therefore convenient to 
define the length $|\cl N|$ of a channel $\cl N$ by the minimum 
number of Kraus operators in an operator-sum representation, or,
equivalently, as the minimum dimension of an ancilla in a Stinespring
representation needed to represent $\cl N$.

According to the above definition a quantum channel maps density
operators to density operators, and therefore should be
trace-preserving. As a matter of fact, it is sometimes advantageous to
be less restrictive and to consider also trace-decreasing channels. 
Being still a completely positive mapping, an in general
trace-decreasing channel $\cl N: 
\cl B(Q) \to \cl B(Q')$ has a Stinespring representation with an
operator $V:Q \to Q'E $ satisfying $V\dag V \le \b 1_Q$. As a
consequence, corresponding Kraus operators $A_1, \dots A_N$ of $\cl N$ may be
incomplete, meaning that $\sum_{k=1}^N A_k\dag A \le \b 1_Q$. 
Physically, a trace-decreasing channel describes a transmission
that involves either some selective process or some leakage, as an 
effect of which a system does not necessarily reach its 
destination. This motivates us to denote $\tr \cl N(\rho) $ as the
transmission probability of state $\rho$ with respect to $\cl 
N$.

\subsection{Fidelities}\label{sub-fidelities}
A frequently used quantity for measuring the distance
of general quantum states is the fidelity \cite{Uhl76,Joz94,NC00}
\be
F(\rho, \sigma) := \tn{\sqrt \rho \sqrt \sigma}^2\:, 
\ee
where $\tn{\dots}$ denotes the trace norm, $\tn{A} = \tr \sqrt{A\dag
  A}$. 
If one of the states is pure, say $\rho = \psi = \pro{\psi}$, this reduces to 
\be
F(\psi, \sigma) = \sp{\psi|\sigma|\psi}\:. 
\ee
Generally, $0 \le F(\rho, \sigma) \le 1$, and $F(\rho, \sigma)=1$ if
and only if $\rho = \sigma$.  
The fidelity of two states is related to their trace norm 
distance by \cite{NC00}
\be
1 - \tn{\rho -\sigma} \: \le \: F(\rho,\sigma) \: \le \: 1 - \fr{1}{4}
\tn{\rho- \sigma}^2\:.
\ee
Furthermore, the fidelity is monotonic under quantum operations in the
sense that for any trace-preserving completely positive $\cl E:\cl
B(Q) \to B(Q')$, 
\be
F(\rho, \sigma) \le F(\cl E (\rho), \cl E(\sigma))\:.
\ee

A remarkably theorem by Uhlmann \cite{Uhl76}  states that the fidelity
of $\rho$ 
and $\sigma$ can be also understood as the maximum transmission
probability $|\sp{ \psi| \fie}|^2$ of purifications $\psi$ and $\fie$
for $\rho$ and $\sigma$, respectively. The fidelity $F(\rho,\sigma)$
thus tells us how close two pure states $\psi$ and $\fie$ of a
universe can be if  
they are known to reduce to states $\rho$ and $\sigma$ on a subsystem
$Q$. More precisely, 
the theorem states that if $\psi_{RQ}$ in $RQ$ is a purification of 
$\rho$,  and if $\sigma$ can be also purified on $RQ$, then 
\be
F(\rho, \sigma) := \max_{\fie_{RQ}}
\: |\sp{\psi_{RQ}|\fie_{RQ}}|^2\:,
\ee
where the maximum is taken over all purifications $\fie_{RQ}$ of
$\sigma$ in $RQ$ \cite{Joz94}. 

To determine how well a state $\rho$ is preserved under a channel 
$\cl E : \cl B(Q) \to \cl B (Q')$ we will generally use 
the entanglement fidelity \cite{Sch96}
\ben\label{Fe-def}
F_e(\rho, \cl E) := \sp{ \psi_{RQ}| \cl I_R \otimes \cl E(\psi_{RQ}) |
  \psi_{RQ}}\:,
\een
where $\psi_{RQ}$ is any purification of $\rho$ on $Q$ extended by an
ancilla system $R$, and $\cl I_R$ is the identity operation on
$R$. In terms of Kraus operators $A_1, \dots, A_{|\cl E|}$ of $\cl E$ the
entanglement fidelity can be expressed as \cite{Sch96}
\ben\label{Fe-sum}
F_e(\rho, \cl E) = \sum_{k=1}^{|\cl E|} |\tr\:\rho A_k|^2\:.
\een
The entanglement fidelity of a state $\rho = \sum_i p_i \psi_i$ 
is known to be a lower bound of the averaged fidelities $F(\psi_i, 
\cl E(\psi_i))$ \cite{NC00}, 
\be
F_e(\rho, \cl E) \: \le \: \sum_i p_i F(\psi_i, \cl E(\psi_i))\:.
\ee
This relation becomes particularly useful if $\rho$ is chosen to be 
the normalized projection $\pi_C$ on a subspace $C$ of $Q$,
$\pi_C = \Pi_C /|C|$. Then the entanglement fidelity yields  a lower
bound of the average subspace fidelity,
\ben\label{average-fidelity}
F_e(\pi_C, \cl E) \:\le \:\int_C d \psi \: F(\psi,\cl E(\psi)) =:
F_{av}(C,\cl E)\:.
\een
where the integral is taken with respect to the normalized, unitarily
invariant
measure on $C$. Actually, there also exists a strict relation between
the two fidelities \cite{HHH99,Nie02}, 
\be
F_{av}(C, \cl E) = \fr{ |C|\: F_e(\pi_C,\cl E) + 1}{|C| +1}\:.
\ee

We emphasize that with Eq.~\Ref{Fe-def} also the entanglement fidelity
with respect to a trace-decreasing channel $\cl E$ is defined. In
this case representation \Ref{Fe-sum} turns out to hold 
as well, leading to the following simple but nevertheless useful
observation. Let  a channel $\cl E: \cl B(Q) \to \cl B(Q')$ be defined
by Kraus operators $A_1, \dots, A_{|\cl E|}$. We call a second channel
$\tilde \cl E: \cl B(Q) 
\to \cl B(Q')$ a reduction of $\cl E$ if it can be represented by a
subset of the Kraus operators $A_1, \dots, A_{|\cl E|}$, i.e.
\be
\tilde \cl E (\rho)= \sum_{k\in \tilde N} A_k \rho A_k\dag\:, \quad
\tilde N \subset \{1, \dots, |\cl E| \}\:.
\ee
By Eq.~\Ref{Fe-sum} we notice that reducing a channel can never
increase entanglement fidelity: for any reduction $\tilde \cl E$ of a
channel $\cl E$
\ben\label{reduced-channel}
F(\rho, \tilde \cl E) \: \le \: F(\rho, \cl E)\:.
\een

\section{Quantum coding theorem}\label{sec-coding}

\subsection{Quantum capacity of a quantum channel}\label{sub-capacity}  %

For the purpose of quantum-information transmission, Alice (sender)
and Bob (receiver) may employ a  quantum channel $\cl N : \cl B(Q) \to
\cl B(Q')$ that conveys an input quantum system $Q$ from Alice to
an in general different output system $Q'$ received by Bob. 

In the simplest case, Alice may prepare quantum information in form of
some state $\rho$ of $Q$, which after transmission via the channel becomes 
a state $\rho'= \cl N(\rho)$ of $Q'$ received by Bob.
In order to obtain Alice's originally sent state $\rho$, Bob may
subject $\rho'$ to suited physical manipulations, which eventually
should result in
a state $\rho''$ of $Q$ close to $\rho$. Mathematically, this corresponds 
to the application of a trace-preserving, completely positive mapping
$\cl R : \cl B(Q') \to 
\cl B(Q)$, which we denote as recovery operation in the following.
Referring to Sec.~\ref{sub-fidelities}, relation \Ref{average-fidelity},
the overall performance of this elementary transmission scheme can be
conveniently assessed by the entanglement fidelity $F_e(\pi, \cl R \circ \cl N)$ of
the homogeneous density $\pi = \b 1_Q/|Q|$ of $Q$ with respect to 
$\cl R \circ \cl N $, or, if we
suppose that Bob has optimized the recovery operation $\cl R$, by
the maximized entanglement fidelity
\ben\label{max}
\max_R F_e(\pi, \cl R \circ \cl N)\:.
\een

To improve the transmission scheme, Alice and Bob may agree upon using
only states $\rho$ 
whose supports lie in a certain linear subspace $C$ of $Q$
\footnote{This can be advantageous when the 
interaction of system and environment does affect 
states in $C$ significantly less than the average state, for instance,
because $C$ obeys certain symmetries of the system-environment
interaction Hamiltonian. Moreover, the restriction to a suited
subspace $C$ may allow Bob to employ quantum error-correcting schemes
in the recovery operation $\cl R$ \cite{Sho95,Ste96}.}.
A subspace used for this purpose 
is called a (quantum) code. Its size $k$ is defined as
$k = \log_2 |C|$, meaning that a pure state in $C$ carries $k$ qubits
of quantum information \cite{Sch95}. Corresponding to \Ref{max}, an
appropriate quantity for assessing the suitability of a code $C$ for a
channel $\cl N$ is the quantity  
\be
F_e(C, \cl N) := \max_{\cl R} F_e(\pi_C, \cl R \circ \cl N)\:,
\ee
where $\pi_C = \Pi_C/|C|$ is the normalized projection on $C$
(again cf.~Sec.~\ref{sub-fidelities}).
We refer to this quantity as the entanglement fidelity of the code $C$ with
respect to the channel $\cl N$. 

The definition involves a non-trivial
optimization of the recovery operation $\cl R$.
At first sight, this makes the code entanglement fidelity rather
difficult to determine and therefore may cast doubts on its
usefulness.
However, following Schumacher and Westmoreland \cite{SW02} we will 
derive a useful explicit lower bound for $F_e(C, \cl N)$ in Sec.~\Ref{sec-lower}.

In the elementary transmission scheme considered so far the quantum
information is encoded in single quantum systems $Q$ and transmitted in single
uses (``shots'') of the channel $\cl N$. Like in classical
communication schemes the restriction to single-shot uses of the
channel is very often far from being optimal. Since the work of
Shannon \cite{SW49} it is known that encoding and transmission of 
information in large blocks yields much better results. 
 
In an $n$-block transmission scheme, Alice uses $n$ identical copies
of the quantum 
system $Q$, in which she encodes quantum information as a state $\rho$
with support 
in a chosen code $C_n \subset Q^n$. 
During the transmission
each individual system $Q$ is independently transformed by the
channel $\cl N$, and Bob receives the state $\cl N\tenpow n(\rho)$,
on which he applies a recovery operation $\cl R_n : \cl B(Q^n) \to \cl
B(Q'^n)$.  
The crucial differences to a single-shot scheme are the usage of a code
$C_n$ and a recovery operation $\cl R_n$ which in general will not
obey the tensor product structure, i.e. 
$C_n \neq C_1\tenpow n$ and $\cl R_n \neq \cl R_1 \tenpow n$.
The rate
$
R = \fr{1}{n} \log_2 |C_n|
$
of an $n$-block code $C_n \subset Q^n$ denotes the average number of
qubits encoded per system $Q$ and sent per channel use. 

In the end, we wish to know up to which rate the channel
$\cl N$ can reliably transmit quantum information 
when an optimal block code $C_n$ of arbitrarily large block number $n$
is used.  This rate defines the { quantum capacity} $Q(\cl N)$ of
the channel $\cl N$ \cite{Llo97,BNS98,BKN00} (for a recent review see
e.g. \cite{KW04}). 
A mathematically precise definition uses the notion of an
{ achievable rate}. A rate $R$ is called achievable by the channel
$\cl N$ if there is a sequence of codes $C_n \subset Q^n$,
$n=1,2,\dots$, such that
\be
\lim_{n\to\infty} {\sup} \fr{\log_2 |C_n|}{n} \ge R\:, \quad \mbox{and}
\quad \lim_{n\to\infty} F_e(C_n,\cl N\tenpow n)  = 1\:.
\ee
The supremum of all achievable rates of a channel $\cl N$ is the
quantum capacity $Q(\cl N)$ of the channel $\cl N$.

\subsection{Quantum coding theorem}\label{sub-coding}
Determining the quantum capacity of a channel $\cl N$ poses 
one of the central problems of quantum information theory. It is
partially solved by the quantum coding theorem
\cite{Llo97,BNS98,BKN00} which relates 
quantum capacity to coherent information \cite{SN96}, the quantum
analogue to mutual 
information in classical information theory.  
The coherent information is defined for a state $\rho$ with respect to
a trace-preserving channel $\cl N$ as
\be
I(\rho, \cl N) \: = \: S(\cl N(\rho)) - S_e(\rho, \cl N)\:.
\ee
This is the von Neuman entropy of the channel output, $S(\cl N(\rho))$,
minus the entropy exchange $S_e(\rho, \cl N)$ between system and
environment, which is given by
\be
S_e(\rho, \cl N) = S(\cl I_R \otimes \cl N(\psi_{RQ}))\:, 
\ee
where $\psi_{RQ}$ is a purification of $\rho$, and $\cl I_R$ is the
identity operation on the ancilla system $R$ \cite{Sch96}. 

The quantum noisy coding theorem states that the quantum capacity
$Q(\cl N)$ of a channel $\cl N$ is the { regularized} coherent
information $I_r(\cl N)$ of $\cl N$, 
\be
Q(\cl N) = I_r(\cl N) :=  \lim_{n\to\infty} \fr{1}{n} \max_{\rho}
I(\rho, \cl N \tenpow n)\:. 
\ee
The limiting procedure corresponds to the one in the definition of 
an achievable rate and thus contributes to the fact that generally
optimal coding can be only asymptotically reached in the limit of 
block numbers $n\to \infty$. As a consequence of this limit the
regularized coherent information and thus the quantum capacity of a
channel is still difficult to determine. 

The regularized coherent information has long been known an upper
bound for $Q(\cl N)$, which is the content of the converse coding
theorem \cite{BNS98,BKN00}. The direct coding theorem, stating that
$I_r(\cl N)$ is 
actually attainable, has been strictly proven first by Devetak
\cite{Dev05}.  
His proof utilizes a correspondence of classical private information
and quantum information. 

Sections \ref{sec-lower}, \ref{sec-average}, \ref{sec-general}, and
\ref{sec-proof} below represent the four stages of 
a different proof for the direct quantum coding theorem, of which an
earlier version appeared in Ref.~\cite{Kle07}.
The working hypothesis underlying this proof is that randomly chosen
block codes of sufficiently large block number 
typically allow for almost perfect quantum error correction. 
In this respect, the present proof as well as the one of Hayden 
{\em et al.} \cite{HHYW07}~and also the earlier approaches of Shor \cite{Sho02}
and Lloyd \cite{Llo97} follow Shannon's original treatment \cite{SW49}
of the classical coding problem. 

\section{Outline of proof}\label{sec-outline}
In the first stage of the the proof (Sec.~\ref{sec-lower}) we
establish a lower bound for the code entanglement 
fidelity. 
It is essentially an earlier result of Schumacher and
Westmoreland \cite{SW02}, of which has been also made good use of
recently by Abeysinghe {\em et al.}\ \cite{ADHW06} and Hayden {\em et
  al.}\ \cite{HHYW07} in the 
same context.  
The bound can be explicitly determined in terms of Kraus
operators of the channel $\cl N$, and its use will relieve us from the
burden of optimizing a recovery operation $\cl R$ for a given code $C$
and channel $\cl N$ in the course of proving the coding theorem. 
In deriving the lower bound the optimization of $\cl R$ is solved by
means of Uhlmann's theorem. 

In the next stage (Sec.~\ref{sec-average}) we investigate the error
correcting 
ability of codes that are chosen at random from a unitarily invariant
ensemble of codes with a given dimension $K$. Taking the average 
of the lower bound derived in Sec.~\ref{sec-lower} we will show 
the averaged code entanglement fidelity of a channel $\cl N:\cl B(Q)
\to B(Q')$ to obey
\ben\label{lb-outline}
\scav{ F_e(C, \cl N)}_K \: \ge \: \tr\:\cl N(\pi) \: - \:\sqrt{ K |\cl
  N|} \fn{ \cl N(\pi)}\:,
\een
where $\pi = \b 1_Q/|Q|$, and $\fn{\dots}$ denotes the Frobenius norm
or two norm.

In Sec.~\ref{sec-unital}
we will illustrate the efficiency of random coding by means of 
the special case of a unital channel $\cl U:\cl B(Q)\to \cl B(Q')$,
which by definition satisfies $\cl U(\pi) 
= \pi'$. In this case the lower bound \Ref{lb-outline} immediately
proves the attainability of the quantum Hamming bound by random
coding, and thus provides evidence for the validity of 
the above mentioned working hypothesis. Moreover, if we demand the
channel $\cl U$ to be also uniform, as will be defined in
Sec.~\ref{sec-unital}, we can 
easily establish the coherent information $I(\pi,\cl U)$ to be a lower bound
of the quantum capacity $Q(\cl U)$, 
\be
Q(\cl U) \: \ge \: I(\pi, \cl U) \:.
\ee

The third stage of the proof (Sec.~\ref{sec-general}) is merely the
generalization of this relation to an arbitrary channel $\cl N:\cl B(Q)
\to \cl B(Q')$. 
To this end we have to consider $n$-block transmission schemes. For large
$n$ it is 
possible to arrange for unitality and uniformity of $\cl N\tenpow n$
in an approximate sense by, as it will turn out, only minor
modifications of $\cl N\tenpow n$. Approximate uniformity is achieved
by reducing the operation $\cl N\tenpow n$ to an operation $\cl N_{\eps,n}$
consisting only of typical Kraus operators. Furthermore, letting
$\cl N_{\eps,n}$ follow a projection on the typical subspace of $\cl N(\pi)$
in ${Q'}^n$ establishes an approximatively uniform and unital
channel $\tilde{\cl N}_{\eps,n}$, which nevertheless is close to the
original $\cl N\tenpow n$. In the end, this suffices to prove 
$Q(\cl N) \ge I(\pi, \cl N)$ for a general channel $\cl N$. 
A corollary is that for any subspace $V \subset Q$ with
normalized projection $\pi_V= \Pi_V/|V|$ 
\be
Q(\cl N) \: \ge \: I(\pi_V, \cl N)\:. 
\ee

Finally, in Sec.~\ref{sec-proof} we employ a lemma of 
Bennett, Shor, Smolin, and Thapliyal (BSST) \cite{BSST02} in order to
deduce from the last relation 
\be
Q(\cl N) \ge \fr{1}{m} I(\rho, \cl N \tenpow m)
\ee
for an arbitrary integer $m$, and any density $\rho$ of $Q^n$. This
shows the regularized coherent information to be a lower bound of 
$Q(\cl N)$ and thus concludes the proof of the direct coding theorem.

\section{A lower bound for the code entanglement fidelity}\label{sec-lower}
Let a (possibly trace-decreasing) quantum channel $\cl N:\cl B(Q) \to
\cl B(Q')$ have a 
Stinespring representation with an operator $V:Q\to Q' E$, 
and let $C \subset Q$ be a code whose normalized projection $\pi_C =
\Pi_C /|C|$ may have a purification $\psi_{RQ}$ on $R Q$, with
$R$ being an appropriate ancilla system.
Following Schumacher and Westmoreland we will establish 
\ben\label{sw}
F_e(C,\cl N) \:\ge\: p - p\tn{\rho_{RE}' - \rho_R \otimes
  \rho_E'}\:,
\een
where $p = \tr \cl N(\pi_C)$, $\rho_R= \tr_Q \:\psi_{RQ},$ 
and the states $\rho_{RE}'$ and  $\rho_E'$ are reduced  
density operators of the final normalized pure state
\ben\label{final}
\psi_{RQ'E}' = \fr{1}{p} (\b 1_R \otimes V) \psi_{RQ} (\b 1_R \otimes V\dag)\:,
\een
\be
\rho_{RE}' = \tr_{Q'}\: \psi_{RQ'E}\:, \quad
\rho_{E}' = \tr_{RQ'} \:\psi_{RQ'E}\:.
\ee
Furthermore, we will show show that the lower bound \Ref{sw} can
alternatively be formulated in terms of Kraus operators $A_1, \dots,
A_N$ of $\cl N$ as  
\ben\label{Fe-lower-bound}
F_e( C, \cl N) \: \ge \: \: p \: - \tn D \:,
\een
where 
\ben\label{Dop}
D = |C| \sum_{ij=1}^N \left( \pi_C A_i \dag A_j \pi_C \: - \: \tr( \pi_C
  A_i\dag A_j \pi_C) \: \pi_C \right) \otimes \ket i \bra j\:,
\een
with $\ket 1, \dots, \ket N$ being orthonormal states of some ancilla
system. 

{\em Proof of relation \Ref{sw}:} We recall that the code entanglement
fidelity involves a non-trivial 
optimization procedure of a recovery operation $\cl R$
(cf.~Sec.~\ref{sub-capacity}). The idea 
is to hand over this job to Uhlmann's theorem. 
To this end we consider the pure state 
\be
\tilde \psi := \psi_{RQ} \otimes \psi_{RQ'E}'
\ee
of the joint system $RSQ'E$, where $S$ denotes a copy of $QR$.
Obviously, $\tilde \psi$ is a purification of the state $\rho_R
\otimes \rho_E'$ with respect to the ancilla $SQ'$.
Next, we extend $\psi_{RQ'E}'$ by the operation 
\be
\cl E:\cl B(Q') \to B(SQ')\:, \: \rho \mapsto \psi_S \otimes \rho\:, 
\ee
where $\psi_S$ is any fixed pure state of $S$, to a pure state
\be
\psi' := \cl I_R \otimes \cl E \otimes \cl I_E (\psi_{RQ'E}')
\ee
of $RSQ'E$. $\psi'$ is a purification of $\rho_{RE}'$ with respect to
$SQ'$, since 
\be
\tr_{SQ'} \psi' = \tr_{Q'} \tr_S \psi' = \tr_{Q'} \psi'_{RQ'E} =
\rho_{RE}'\:.
\ee
Now, let another purification $\fie$ of $\rho_{RE}'$ in $RSQ'E$
maximize the transition amplitude to $\tilde \psi$, 
\be
|\sp{ \tilde \psi | \fie }|^2 = 
\max_{
 \mbox{ $\chi$ {\scriptsize purification of
    } $\!\rho_{RE}'$}} 
|\sp{ \tilde \psi | \chi }|^2\:.
\ee
According to Uhlmann's theorem (cf.~Sec.~\ref{sub-fidelities}) we know that
\ben\label{scalar-product}
|\sp{ \tilde \psi | \fie }|^2 
= F(\rho_R \otimes \rho_E', \rho_{RE}')\:.
\een
Then, an optimal recovery operation $\cl R : \cl B(Q') \to \cl B(Q)$ can be
constructed my means of a unitary operation $U_{SQ'}$ on $SQ'$ that
rotates the actual (extended) final state $\psi'$ to the maximizing
state $\fie$, 
\be
\fie = (\b 1_R \otimes U_{SQ'} \otimes \b 1_E) \psi' (\b 1_R \otimes
U_{SQ'}\dag \otimes \b 1_E)\: .
\ee
Keeping in mind that $S=QR$ we define 
\be
\cl R(\rho_{Q'}) := \tr_{RQ'} U_{SQ'} \cl E(\rho_{Q'}) U_{SQ'}\dag\:,
\ee
and realize that for the state $\rho_{RQ'} = \tr_{E}\psi'_{RQ'E}$
\bea
\cl I_R \otimes \cl R(\rho_{RQ'}') 
&=&
\tr_{RQ'} \:
(\b 1_R \otimes U_{SQ'}) \cl I_R \otimes \cl E (\rho_{RQ'}') (\b 1_R
\otimes U_{SQ'}\dag) \\
&=& 
\tr_{RQ'E} \:
(\b 1_R \otimes U_{SQ'} \otimes \b 1_E)
\psi' (\b 1_R \otimes U_{SQ'}\dag \otimes \b 1_E) \\
&=& 
\tr_{RQ'E} \:  \fie\:, 
\eea
where here and in the following the partial trace over $R$ refers to
the second $R$ appearing in the product Hilbert space $RSQ'E =
RQRQ'E$. 
Since further 
\be
\psi_{RQ} = \tr_{RQ'E} \tilde \psi\:, 
\ee
we conclude 
\bea
F_e(\pi_C, \cl R \circ \cl N) &\ge &
p\: F(\psi_{RQ}, \cl I_R \otimes \cl R(\rho_{RQ'}')) \\
&=& p\: F(\tr_{RQ'E}\: \tilde \psi, \tr_{RQ'E}\: \fie) \\
&\ge & 
p|\sp{\tilde \psi|  \fie}|^2\:,
\eea
where the second inequality is due to the monotonicity of the fidelity
under partial trace. 
With Eq.~\Ref{scalar-product} and the general relation $F(\rho,
\sigma) \ge 1 - \tn{\rho-\sigma}$ this proves relation \Ref{sw}.

{\em Proof of relation \Ref{Fe-lower-bound}:} 
We choose a purification $\psi_{RQ}$ of $\pi_C$ 
with a state vector
\be
\ket \psi_{RQ} = \fr{1}{\sqrt{ K}} \sum_{l=1}^{K}
\ket{c_l^R}\ket{c_l^Q}\:,
\ee
where $K=|C|$, and $\ket{c_1^R}, \dots \ket{c_{K}^R}$ and
$\ket{c_1^Q}, \dots \ket{c_{K}^Q}$ denote orthonormal vectors 
that span $R$ and $C$, respectively. 
Supposing that the orthonormal states $\ket 1, \dots, \ket N$ 
span the ancilla $E$ and the Kraus operators $A_1,\dots A_N$ are
associated to $V$ by $A_i \ket   \psi_Q = \sp{i|V|\psi_Q}$,
we immediately obtain from Eq.~\Ref{final} 
\be
p \:\psi_{RQ'E}' = \fr{1}{K} \sum_{lm=1}^K\sum_{ij=1}^N 
\ketbra{c_l^R}{c_m^R}
\otimes A_i
\ketbra{c_l^Q}{c_m^Q}
A_j\dag 
\otimes
\ketbra{i}{j}\:.
\ee
Hence
\bea
p\:\rho_{RE}' &=& \fr{1}{K} \sum_{lm=1}^K \sum_{ij=1}^N
\sp{c_m^Q|A_j\dag A_i | c_l^Q}\:
\ketbra{c_l^R}{c_m^R}
\otimes
\ketbra{i}{j}\\
p\: \rho_R \otimes \rho_E' 
&=& \fr{1}{K^2} \sum_{m=1}^K
\ketbra{c_m^R}{c_m^R}
\otimes
\sum_{l=1}^K
\sum_{ij=1}^N
\sp{c_l^Q|A_j\dag A_i | c_l^Q}\:
\ketbra{i}{j}\:.
\eea
The trace norm of $p(\rho_{RE}'- \rho_R \otimes \rho_E')$ appearing in the
lower bound \Ref{sw} becomes more handy if we transform the
operator difference by an isometry $\cl J:\cl B(RE) \to \cl B(QE)$, 
\be
\cl J: 
\sum_{lm,ij}\alpha_{lm,ij} 
\ketbra{c_l^R}{c_m^R}
\otimes
\ketbra{i}{j}\:
\mapsto
\:
\sum_{lm,ij}\alpha_{lm,ij}^*
\ketbra{c_l^Q}{c_m^Q}
\otimes
\ketbra{i}{j}\:.
\ee
$\cl J$ shifts from $R$ to $Q$ and then complex conjugates with
respect to the basis $\ket{ c_l^Q} \otimes \ket{ i }$, which 
clearly leaves the trace norm invariant. 
A straightforward calculation then shows
\be
D 
:=
p \cl J(\rho_{RE}'- \rho_R \otimes \rho_E') 
\: = \:
 K
\sum_{ij=1}^N 
\left(
\pi_C A_i\dag A_j \pi_C \: - \: \tr(\pi_C A_i\dag A_j \pi_C)\: \pi_C
\: 
\right) 
\otimes
\ketbra{i}{j}\:,
\ee
as in Eq.~\Ref{Dop}, and further 
\be
F_e(\cl C, \cl N) 
\: \ge \:
p\: - p \tn{\rho_{RE}'- \rho_R \otimes \rho_E' } 
\: = \:
p \: - \tn{ \cl J(\rho_{RE}'- \rho_R \otimes \rho_E' )} 
\: = \: p\: - \tn D\:.
\ee
which is what we wanted to proof. 

\section{Random coding}\label{sec-average}
Let the unitarily invariant code ensemble of all $K$-dimensional
codes $C \subset Q$ be defined by the ensemble average
\ben\label{average}
\scav{ A(C) }_K \: := \: \int_{\b U(Q)} d \mu(U) \: A(U C_0)
\een
of a code dependent variable $A(C)$. Here, $C_0$ is some fixed
$K$-dimensional code space in $Q$, and $\mu$ is the normalized Haar
measure on $\b U(Q)$, 
the group of all unitaries on $Q$. 
Below we will show that the ensemble averaged code entanglement
fidelity of a (possibly trace-decreasing) channel $\cl N:\cl B(Q) \to
\cl B(Q')$ obeys 
\ben\label{Fe-lb-N}
\scav{ F_e(C, \cl N) }_K \: \ge \: \tr\:\cl N(\pi) \: - \: \sqrt{K|\cl
  N|} \fn{ \cl N(\pi)}\:,
\een
where $\pi=\b 1_Q/|Q|$ is the uniform density on $Q$.

We begin with the ensemble average of relation \Ref{Fe-lower-bound}, 
\ben\label{average-Fe}
\scav{F_e(C,\cl N)}_K \: \ge \: \scav{ \tr \cl N(\pi_C)}_K - \scav{ \tn D }_K\:, 
\een
where, as always, $\pi_C= \Pi_C/|C|$, and 
\ben\label{Dop2}
D = K \sum_{i,j=1}^{N} \left(
\pi_C A_i\dag A_j \pi_C \: - \: \tr(\pi_C A_i\dag A_j \pi_C)\: \pi_C
\: 
\right) 
\otimes
\ketbra{i}{j}\:,
\een
with  $A_1,\dots, A_N$ being $N=|\cl N|$  Kraus operators of a minimal
operator-sum representation of $\cl N$.   To average $\tr \cl
N(\pi_C)$ we realize that $\rho \mapsto \tr \cl N(\rho)$ as a linear
operation interchanges with the average. 
Since $\scav{ \pi_C }_K = \pi$ we thus obtain
\be
\scav{  \tr \cl N(\pi_C)  }_K = \tr \cl N(\scav {\pi_C}_K) = \tr \cl N(\pi)\:.
\ee
Directly averaging the trace norm of $D$ turns out to be quite
cumbersome. Therefore, we first estimate
\be
\scav{ \tn D  }_K^2 \: \le \: K N \scav{ \fn D }_K^2 \: \le \: KN \scav{
  \fn{ D }^2 }_K\:,
\ee
where $\fn D \: = \: (\tr\: D\dag D)^{1/2}$ denotes the Frobenius norm
(two-norm) of $D$. The first inequality follows from the general
relation $\tn A \le \sqrt{d} \fn A$, where $d$ is the rank of $A$, 
and the second inequality is Jensen's inequality. 
This leads us to 
\ben\label{Fe-lb-D}
\scav{F_e(C,\cl N)}_K \: \ge \: \tr \cl N(\pi)\: - \: \sqrt{ KN\: \scav{
    \fn{ D }^2 }}_K\:, 
\een
and it remains to determine the ensemble average of $\fn{ D }^2$.
From the explicit representation Eq.~\Ref{Dop2} follows 
\be
\fn{ D }^2 = \tr \: D\dag D \: = \:
\sum_{ij=1}^N 
\tr( \pi_C W_{ij}\dag \pi_C W_{ij}) 
- \fr{1}{K} |\tr \: \pi_C
W_{ij}|^2\:,
\ee
where operators $W_{ij}$ are
\be
W_{ij}  = A_i\dag A_j\:.
\ee
It is useful to introduce a Hermitian form
\ben\label{b-def}
b(V,W)\: := \: \scav{
\tr ( \pi_C V\dag \pi_C W)  \: - \:
\fr{1}{K} \tr( \pi_C V\dag) \: \tr( \pi_C W)
}_K\:,
\een
with which
\ben\label{b-average}
\scav{ \fn{ D}^2 }_K  \: = \: \sum_{ij=1}^N b(W_{ij}, W_{ij})\:.
\een
The point is that the unitary invariance of the ensemble average
entails the unitary invariance of $b$, i.e., for any $U\in \b U(Q)$   
\be
b(V,W) \: = \: b(UVU\dag, U W U\dag)\:.
\ee
which, in fact, already determines $b$ to a large extend:
According to Weyl's theory of group invariants \cite{Wey46,How92}
$b(V,W)$ must be a linear 
combination of the only two fundamental 
unitarily invariant Hermitian forms $\tr\: V\dag W$ and 
$\tr\: V\dag \: \tr\: 
W$, 
\ben\label{lc}
b(V,W) \: = \: \alpha \: \tr\: V\dag W\: + \beta \: \tr\: V\dag \: \tr
\: W\:.
\een
An elementary proof of this fact is outlined in Appendix
\ref{app-unitary-invariance}.
To determine the coefficients $\alpha$ and $\beta$ we consider two
special choices of the operators $V$ and $W$. For $V=W=\b 1_Q$
Eqs.~\Ref{b-def} and \Ref{lc} yield 
\ben\label{I}
\alpha M + \beta M^2 = \fr{1}{K}\:,
\een
where here and henceforth $M=|Q|$.
Secondly, when we set $V$ and $W$ to a projection $\psi =
\ketbra{\psi}{\psi} $ on $Q$ we obtain from Eq.~\Ref{b-def}
\be
b(\psi,\psi) = \fr{K-1}{K} \scav{ |\sp{ \psi|\pi_C|\psi }|^2 }_K\:.
\ee
Reverting to random matrix theory we find in Appendix \Ref{app-rmt}
$ \scav{ |\sp{ \psi|\pi_C|\psi }|^2 }_K = (1+1/K)/(M^2+ M) $, and hence
\be
b(\psi,\psi) = \fr{1-K^{-2}}{M^2 +M}\:.
\ee
With $b(\psi,\psi)= \alpha + \beta $ from Eq.~\Ref{lc} this yields the
second equation,
\ben\label{II}
\alpha + \beta = \fr{1-K^{-2}}{M^2 +M}\:.
\een
Solving Eqs.~\Ref{I} and \Ref{II} for $\alpha$ and $\beta$, and
inserting the solution into \Ref{lc} produces
\be
b(V,W) = \fr{1-K^{-2}}{M^2-1} \left( \tr\:V\dag W \: 
- \: \fr{1}{M} \tr\:V\dag \: \tr\:W \right)\:,
\ee
and, by Eq.~\Ref{b-average}, 
\ben\label{exact-average}
\scav{ \fn{ D }^2 }_{K} =\fr{1-K^{-2}}{M^2-1} \sum_{ij} 
\left(\tr\:W_{ij}\dag W_{ij}  
- \fr{1}{M} |\tr\:W_{ij}|^2  \right)\:.
\een
In general, not much is given away 
when we use the upper bound for $\scav{ \fn{ D }^2}_{K}$ that we obtain by using
$(1-1/K^2)/(M^2-1) \le 1/M^2$ and by omitting the negative terms
$-|\tr\:W_{ij}|^2/M $ in the sum. Then
\be
\scav{ \fn{ D }^2 }_{K} \: \le \: \fr{1}{M^2} \sum_{ij} 
\tr\:W_{ij}\dag W_{ij} 
= \tr(\sum_j A_j \fr{\b 1_Q}{M} A_j\dag \sum_i A_i \fr{\b 1_Q}{M} A_i\dag
)\:,
\ee
where we cyclically permuted operators under the trace to obtain the
last equality. We realize that the argument of the trace is simply $\cl
N(\pi)^2$  (with $\pi = \b 1_Q/M$). 
This yields the rather simple upper bound 
\ben\label{final-D-upperbound}
\scav{ \fn{ D }^2 }_{K} \: \le \:   \fn{ \cl N( \pi)}^2\:,
\een 
which finally proves the lower bound \Ref{Fe-lb-N}  by relation
\Ref{Fe-lb-D}.

\section{Unital and uniform channels}\label{sec-unital}
The efficiency of random coding can be easily demonstrated by relation 
\Ref{Fe-lb-N} for the case of a { unital} channel 
$\cl U:\cl B(Q) \to \cl B(Q')$, which by definition
maps the homogeneously distributed input state $\pi$ to the homogeneously
distributed output state $\pi'$.
An example is a random unitary channel $\cl U_r: \cl B(Q) \to \cl
B(Q)\:, \rho \mapsto \sum_i p_i U_i \rho U_i\dag$, where arbitrary
unitary operators $U_1, \dots, U_N$ are applied with probabilities
$p_1, \dots, p_N$ on the system $Q$.

Thus, for a  unital channel $\fn{ \cl U(  \pi) } = \fn{
  \pi'} = |Q'|^{-1/2}$, which by relation \Ref{Fe-lb-N} predicts the average
entanglement fidelity of $K$-dimensional codes to obey
\be
\cav{ F_e(C, \cl U)}_{K} \: \ge \: 1 - \sqrt{\fr{K |\cl U|}{|Q'|}} \:.
\ee
This means that almost all codes of dimension $K$ allow for almost
perfect correction of the unital noise $\cl U$, provided that
\be 
K |\cl U| \ll |Q'|\:.
\ee
Recalling that $|\cl U|$ is the number of Kraus operators in an
operator-sum representation of $\cl U$, this relation
clearly shows 
the attainability of the quantum Hamming bound \cite{EM96} by random
coding. Formally, this is equivalent to the 
lower bound 
\ben\label{Q-of-U}
Q(\cl U) \: \ge \: \log_2 |Q'| - \log_2 |\cl U|
\een
of the quantum information capacity of $\cl U$. To see this, we 
consider the $n$-times replicated noise $\cl U\tenpow n$,
and study the averaged  entanglement fidelity of codes
with dimension $K_n = \lfloor 2^{nR} \rfloor$ for some positive rate $R$.
Since with $\cl U$ also $\cl U\tenpow n$ unital, and $|\cl U\tenpow n|
= |\cl U|^n$, this time we arrive at
\be
\cav{ F_e(C, \cl U\tenpow n)}_{K_n} \: \ge \: 1 - \left(
  \fr{2^{R}\: |\cl U|}{|Q'|}   \right)^{n/2}\:. 
\ee
For $n\to \infty$ the right hand side converges to unity if 
$R < \log_2 |Q'| - \log_2 |\cl U|$. 
Hence, all rates below $\log_2 |Q'|-\log_2 |\cl U|$ are achievable, which
by the definition of quantum capacity (cf.~Sec.~\ref{sub-capacity})
shows relation \Ref{Q-of-U}. 

Finally, let us assume that the 
channel $\cl U$ is also
uniform, meaning that $\cl U$ has a minimal operator-sum
representation with Kraus operators $A_1, \dots, A_{|\cl U|}$ obeying 
$\tr A_i\dag A_j = 0 $ for $i\neq j$ and 
$\fr{1}{|Q|} \tr A_i\dag A_i= const. = |\cl U|^{-1} $. 
The first condition is actually no restriction, because a non-diagonal
representation can always be transformed to a diagonal
one\footnote{\label{foot} For arbitrary operation elements $B_1, 
  \dots, B_N$ of $\cl N$, $N = |\cl N|$, let an $N\times N$ matrix $H$
  be defined by $
H_{ij} := \tr B_i\dag B_j\:.
$
Since $H=H\dag$, there is a unitary matrix $U$ such that $U H U\dag$
is diagonal. Because of the unitary freedom in the operator-sum
representation \cite{NC00},
the operators $A_m:= \sum_j U\dag_{jm} B_j$ equivalently represent $\cl
N$. It is readily verified that $\tr A_l\dag A_m = 0$ for $l\neq m$.}
. The second 
condition demands that errors $E_i$ associated with Kraus operators $A_i$
appear with equal probability $p_i=1/|\cl U|$.
We observe that by Schumacher's representation \cite{Sch96} the entropy 
exchange of $\pi$ under a uniform $U$ is simply given by
\be 
S_e(\pi, \cl U)
= S( \b 1_{|\cl U|} / |\cl U|) = \log_2 |\cl U|\:.
\ee
Since $\cl U$ is unital we also have 
\be 
S(\cl U(\pi)) = S(\pi') = \log_2 |Q'|\:. 
\ee
Comparing these expressions with relation \Ref{Q-of-U} and
recalling the definition of coherent information
(cf. Sec.~\ref{sub-coding}) establishes the lower bound 
\be
Q(\cl U) \ge I(\pi,\cl U)\:.
\ee
In fact, the following section we will show this bound to hold
for general channels.

\section{General channels}\label{sec-general}
Starting again with relation \Ref{Fe-lb-N} we will proof for a
general channel $\cl N:\cl B(Q) \to \cl B(Q')$
\ben\label{Q>I}
Q(\cl N) \: \ge \: I(\pi, \cl N)\:,
\een
where $\pi = \b 1_Q/|Q| $, and, as a corollary, 
\ben\label{Q>IV}
Q(\cl N) \: \ge \: I(\pi_V, \cl N)\:,
\een
where $\pi_V$ is the normalized projection $\pi_V = \Pi_V/|V|$ on any
subspace $V\subset Q$. 

The strategy of proving is to approximate $\cl N\tenpow n$ by an almost 
uniform and unital channel $\tilde \cl N_{\eps,n}$, with which we then
proceed as in the preceding section. 
We construct $\tilde \cl N_{\eps,n}$ in two steps. The first step is to
reduce $\cl N\tenpow n$ to its { typical Kraus operators}, as will
be defined below. This yields an almost uniform operation $\cl
N_{\eps,n}$. In a second step, we let $\cl N_{\eps, n}$ follow a
projection on the typical subspace of $\cl N(\pi)$ in ${Q'}^n$, resulting
in an operation $\tilde \cl N_{\eps,n}$ with the 
desired properties. 

We begin with briefly recalling definitions and basic properties
of both typical sequences \cite{CT91} and typical subspaces
\cite{Sch95,NC00}.

\subsection{Typical sequences}\label{subsubsec-typical_sequences}
Let $X_1,\: X_2, \: X_3, \: \dots$ be independent random variables
with an identical probability distribution $\cl P$ over an alphabet
$\aleph$.
Let   
$
H(\cl P) = - \sum_{a\in \aleph} \cl P(a) \log_2 \cl P(a)
$
denote the Shannon entropy of $\cl P$, let $n$ be a positive integer,
and let $\eps$ be some positive number. A sequence $\b a = (a_{1}, a_{2}, 
\dots, a_{n}) \in \aleph^n $ is defined to be $\eps$-typical if its probability 
of appearance 
$
p_{\b a} = \cl P(a_1) \cl P(a_2) \dots \cl P(a_n)
$
satisfies 
\be
   2^{- n( H(\cl P) + \eps)} \: \le \: p_{\b a} \: \le \: 2^{-n( H(\cl
      P) - \eps)}\:.
\ee
Let $\aleph_{\eps,n}$ denote the set of all $\eps$-typical sequences of length $n$.
 
Below we will make use of the following two well-known facts:
\begin{description}

\item[(i)] The number $|\aleph_{\eps,n}|$ of all $\eps$-typical sequences of
  length $n$ is less than $2^{n(H(\cl P) +\eps)}$.

\item[(ii)] The probability $P_{\eps,n} = \sum_{\b a \in  \aleph_{\eps,n}}
   p_{\b a} $ of a random sequence of length $n$ 
being $\eps$-typical exceeds $ 1- 2 e^{-n \psi(\eps)}$, 
where $\psi(\eps)$ is a positive number independent of $n$.
\end{description}
Proofs can be found in Appendix \ref{app-typical}.

\subsection{Typical subspaces}\label{subsub-typical_subspace}
Let $\rho$ be some density operator of a quantum system $Q$, let $n$
be a positive integer, and let $\eps$ be a positive number. 
An eigenvector $\ket{\b v}$ of $\rho\tenpow n$ is called $\eps$-typical if
its eigenvalue $p_{\b v}$ satisfies
\be
2^{-n( S(\rho) + \eps)} \: \le \: p_{\b v} \: \le \: 
2^{-n( S(\rho) - \eps)}\:.
\ee
The $\eps$-typical subspace $T_{\eps,n}$ of $\rho$ in $Q\tenpow n$ is
defined as the span of all $\eps$-typical eigenvectors of $\rho\tenpow
n$. We denote the projection on $T_{\eps,n}$ by $\Pi_{\eps,n}$.

Notice that typical eigenvectors correspond to typical sequences 
when an orthonormal eigen-system $\ket{v_1}, \dots, \ket{ v_{|Q|}} $ 
of $\rho$ is chosen as alphabet $\aleph$, a sequence of length $n$
over $\aleph$ is identified with an eigenvector $\ket{\b v} = 
\ket{v_{j_1}} \ket{v_{j_2}} \dots \ket{v_{j_n}}$ of $\rho \tenpow n$,
and the probability $\cl P(\ket{\b v})$ of an eigenvector $\ket{ \b v}$ of
$\rho$ is taken to be its eigenvalue.
Then, the above stated properties of typical sequences translate to 
 
\begin{description}

\item[(i')] The dimension of $T_{\eps,n}$ is less than $2^{n( S(\rho) +
    \eps)}\:.$

\item[(ii')] The probability $P_{\eps,n} = \tr \: \Pi_{\eps,n} \rho \tenpow n$
  of measuring an $\eps$-typical eigenvalue of $\rho\tenpow n$
  exceeds $1- 2 e^{-n \psi(\eps)}$, 
where $\psi(\eps)$ is a positive number independent of $n$.

\end{description}

\subsection{Reduction of  $\cl N \tenpow n$}\label{subsub-reduction}

Let a trace-preserving channel $\cl N : \cl B(Q) \to \cl B(Q')$ be
given. $\cl N$ may be represented in a minimal operator sum  
with Kraus operators $A_1, \dots, A_{|\cl N|}$, which without loss of
generality we assume to be diagonal, i.e.~$\tr A_j\dag A_i =0$
for $i\neq j$ (cf.~footnote \ref{foot}).
Accordingly, $\cl N\tenpow n$ can be represented by $|\cl N|^n$ Kraus
operators
$ 
A_{j_1} \otimes A_{j_2} \otimes \dots \otimes A_{j_n}
$ 
where $j_\nu = 1, \dots, |\cl N|$. 

Now, letting an alphabet $\aleph$ be defined as the set of Kraus operators $A_1,
\dots, A_{|\cl N|}$ of $\cl N$, the Kraus operators of $\cl N\tenpow
n$ can obviously be regarded as sequences over $\aleph$ of
length $n$. In order to identify an $\eps$-typical sequence of 
length $n$, and with it also an $\eps$-typical Kraus operator of $\cl
N\tenpow n$, we define a probability distribution $\cl P$ over
$\aleph$
by
\be
\cl P (A) = \fr{1}{|Q|} \tr \: A\dag A\:, \quad \mbox{} \quad A
\in \aleph\:.
\ee
The normalization of $\cl P$ follows from the completeness relation
$\sum_{A\in \aleph} A\dag A = \b 1_{Q}$, and, owing to the diagonality
of the Kraus operators, the Shannon entropy
$H(\cl P)$ turns out to agree with the entropy exchange $S_e(\pi,\cl
N)$: Again by Schumacher's representation \cite{Sch96}, 
\be
S_e(\pi, \cl N) = S\left(\sum_{i=1}^{|\cl N|} \fr{1}{|Q|}
\tr(A_i\dag A_i) \ket{i}\bra{i}\right) = S\left( \sum_{i=1}^{|\cl N|}
\cl P (A_i)
\ket i \bra i \right) = H(\cl P)\:.
\ee

Being in the possession of the probability distribution $\cl P$ over the set of
Kraus operators $\aleph$, we can define the $\eps$-typical channel $\cl N_{\eps,n}$
of $\cl N\tenpow n$ to consist precisely of the 
operators $\b A$ that are $\eps$-typical with respect to
$\cl P$, 
\be
\rho \mapsto \cl N_{\eps,n} (\rho) := \sum_{\b A \in \aleph_{\eps,n}}
\b A \rho \b A\dag\:.
\ee
As a direct consequence of properties (i) and (ii) of typical
sequences one finds (cf.~Appendix \ref{app-properties})
\bea
|\cl N_{\eps,n}| &\le & 2^{n( S_e(\pi, \cl N) + \eps)}\:, \\
\tr\: \cl N_{\eps,n}(\pi_n) 
& \ge & 1 - 2 e^{n \psi_1(\eps)}\:, 
\eea
where $\pi_n = \b 1_{Q^n}/|Q|^n$, and $\psi_1(\eps)$ is a positive
number independent of $n$. Furthermore, the relative weight $
\fr{1}{|Q|^n} \tr\b A\dag \b A$ of 
an $\eps$-typical operator $\b A = A_{j_1}\otimes \dots \otimes A_{j_n}$ is
just the probability  $p_{\b A}= \cl P(A_{j_1}) \dots \cl P(A_{j_n})$ and therefore
obeys
\be 
2^{-n(S_e(\pi, \cl N) +\eps)} \le p_{\b A} \le 2^{-n(S_e(\pi, \cl N) -\eps)}\:.
\ee

Hence, keeping only  the $\eps$-typical Kraus operators the original
channel $\cl N\tenpow n$ reduces to a channel $\cl N_{\eps, n}$ with
Kraus operators $\b A \in \aleph_{\eps,n}$ of similar probability
$p_{\b A}$. In general, this strongly reduced the number of Kraus
operators from $|\cl N|^n$ to $|\cl N_{\eps,n}|$ 
and renders $\cl N_{\eps,n}$ much closer to a uniform channel
than the original channel $\cl N\tenpow n$. At the same time, 
the transmission probability of the homogeneously mixed state $\pi_n$
deviates  only by an exponentially small amount from unity. 

In order to achieve also approximate unitality, we will 
further modify the channel by 
letting $\cl N_{\eps,n}$ follow a 
projection $\cl T_{\eps,n} :\rho \mapsto \Pi_{\eps,n} \rho \:\Pi_{\eps,n}$
on the $\eps$-typical subspace  $T_{\eps,n} \subset
Q^n$ of the density $\cl N(\pi)$. This defines the { $\eps$-reduced
operation of $\cl N\tenpow n$} by 
\be
\tilde{\cl N}_{\eps,n} := \cl T_{\eps,n} \circ \cl N_{\eps,n}\:,
\ee
with the following properties shown in Appendix \ref{app-properties}:
\bean
\label{N-eps}
|\tilde \cl N_{\eps,n}| &\: \le \: & 2^{n(  S_e(\pi,\cl N) + \eps)}\:, \\
\label{trace-eps}
\tr\: \tilde{\cl N}_{\eps,n} ( \pi_n )  
&\: \ge \: &
1 - 4 e^{-n \psi_3(\eps)}\:, \\
\label{clN-eps}
\fn{ \tilde{\cl N}_{\eps,n} (\pi_n) }^2  
&\: \le \: &
 2^{-n ( S( \cl N(\pi)) - 3 \eps)}\:, 
\eean
where $\psi_3(\eps)$ is a positive number independent of $n$.
Now we are ready to proof relation \Ref{Q>I}:

\subsection{$Q(\cl N) \ge I(\pi,\cl N)$}\label{subsubsec-lower_boundsQ} 
We note that for any code $C \subset Q\tenpow n$ 
\ben\label{Fe-inequality}
F_e(C, \cl N\tenpow n) 
\: \ge \: 
F_e(C, \cl N_{\eps,n}) 
\: \ge \:
F_e(C, \tilde{\cl N}_{\eps,n})\:.
\een
The first inequality holds because $\cl N_{\eps,n}$ is a reduction of
$\cl N\tenpow n$ (cf.\ Sec.~\ref{sub-fidelities}, relation \Ref{reduced-channel}),
and the second one follows from
\be
\max_{\cl R} F(\pi_C, \cl R \circ \cl N_{\eps,n}) 
\: \ge \: 
\max_{\cl R} F(\pi_C, \cl R \circ \cl T_{\eps,n} \circ \cl N_{\eps,n}) 
\: = \: 
\max_{\cl R} F(\pi_C, \cl R \circ \tilde \cl N_{\eps,n}) \:.
\ee
Averaging relation \Ref{Fe-inequality} over the unitary ensemble
of codes $C \subset Q^n$ of dimension 
\be
K_n = \lfloor 2^{n R} \rfloor\:,
\ee
we immediately obtain with relation \Ref{Fe-lb-N} and the bounds
\Ref{N-eps}, \Ref{trace-eps}, \Ref{clN-eps}
\bea
\scav{F_e(C, \cl N\tenpow n)}_{K_n}
& \ge & \tr \tilde \cl N_{\eps,n}(\pi_n) - \sqrt{K_n |\tilde \cl
  N_{\eps,n}|} \fn{ \tilde \cl N_{\eps,n}(\pi_n)} \\[0.3cm]
&\ge &
1 - 4 e^{-n \psi_3(\eps)} -
2^{\fr{n}{2} \left(
R + S_e(\pi, \cl N) - S( \cl N(\pi)) + 4 \eps \right)}\:.
\eea
For all $\eps > 0$, the right-hand side of inequality
converges to unity in the limit $n \to \infty$ if
the asymptotic rate $R$ obeys 
\be
 R + 4 \eps \: < \: S(\cl N(\pi)) - S_e(\pi, \cl N) \equiv I(\pi, \cl N)\:.
\ee
That is, all rates $R= \lim_{n \to \infty} \fr{1}{n} \log_2 K_n$ below
$I(\pi,\cl N)$ are achievable and 
therefore $I(\pi,\cl N)$ is a lower bound of the capacity $Q(\cl N)$. 

Relation \Ref{Q>IV} follows as a corollary: 

\subsection{$Q(\cl N) \ge  I(\pi_V,\cl N)$ }\label{subsubsec-lower_boundsQ_V} 
Let $V$ be any subspace of the input Hilbert space $Q$ of a channel
$\cl N : B(Q) \to B(Q')$, and let $\pi_V= \Pi_V/|V|$ be the normalized
projection on $V$. 
The restriction of $\cl N$ to densities with support in $V$, 
\be
\cl N_V: B(V) \to B(Q')\:, \: \rho \mapsto \cl N(\rho)\:,
\ee
is a channel for which the result of the previous subsection
obviously predicts $I(\pi_V, \cl N_V)$ an achievable
rate. 
It is evident that then $I(\pi_V, \cl N) = I(\pi_V,\cl N_V)$ is also
an achievable rate 
of the complete channel $\cl N$. Thus, for any subspace $V \subset Q$
\be
Q(\cl N) \ge I(\pi_V, \cl N)\:.
\ee

\section{$Q(\cl N) \ge I_r(\cl N)  $}\label{sec-proof}
Finally, we will show that with the BSST lemma 
the result of the last subsection implies the lower bound
\be
Q(\cl N) \ge \fr{1}{m} I(\rho, \cl N\tenpow m)\:,
\ee
where $m$ is an arbitrary large integer, and $\rho$ any density on
$Q^m$. Clearly, this suffices to prove the regularized
coherent information $I_r(\cl N)$ (cf.\ Sec.\ \ref{sub-coding} )
a lower bound of $Q(\cl N)$.  

The BSST lemma \cite{BSST02} states that for a channel $\cl N$ and an
arbitrary state $\rho$ on the input space of $\cl N$ 
\be
\lim_{\eps \to 0} \lim_{n \to \infty} \fr{1}{n} S(\cl N\tenpow n(
\pi_{\eps,n}) ) \: = \: S(\cl N(\rho))\:, 
\ee
where $\pi_{\eps,n}$ is the normalized projection on the
frequency-typical subspace $T^{(f)}_{\eps,n}$ of $\rho$.  
As a corollary, one obtains an analogous relation for
the coherent information, 
\be
\lim_{\eps \to 0} \lim_{n \to \infty} \fr{1}{n} I(\pi_{\eps,n}, \cl
N\tenpow n) 
 \: = \: I( \rho, \cl N )\:. 
\ee
$T^{(f)}_{\eps,n}$ is similar to the ordinary typical
subspace $T_{\eps,n}$ which we have used above. The difference is
that for $T^{(f)}_{\eps,n}$ typicality of a sequence is defined via 
the relative frequency of symbols in this sequence, whereas for
$T_{\eps,n}$
it is defined by its total probability. 
For details we refer the reader to the work of Holevo \cite{Hol02},
where an elegant proof of the BSST lemma is given. 

Here, what matters is solely the fact that $\pi_{\eps,n}$ is a 
homogeneously distributed subspace density of the kind that we
used in the previous subsection.  
Thus we can make use of the bound $Q(\cl E) \ge I(\pi_V, \cl E)$ 
with, for instance, $\cl E = \cl N \tenpow {mn}$, and $V$ being   
the frequency-typical subspace $T_{\eps,n}^{(f)} \subset Q^{mn}$ of an
arbitrary density $\rho$ on $Q^m$. This means 
that for any  
$\eps > 0$ and any $m,n$ 
\be
Q( \cl N \tenpow {mn} ) \ge I( \pi_{\eps,n}, \cl N \tenpow {mn})\:.
\ee
Using the trivial identity $Q(\cl N\tenpow k) = k Q(\cl N)$ we can
therefore write  
\bea
Q(\cl N) &=& \fr{1}{m} \lim_{n\to\infty} \fr{1}{n} Q(\cl
N\tenpow{mn})\\
& \ge  &
\fr{1}{m} \lim_{\eps \to 0} \lim_{n \to \infty} 
  \fr{1}{n} I( \pi_{\eps,n}, (\cl N \tenpow m) \tenpow n) \\
& =& 
 \fr{1}{m} I(\rho, \cl N\tenpow m)\:,
\eea
where the last equation follows from the corollary.
\vspace{1cm}

I would like to thank Michal Horodecki and Milosz Michalski for
inviting me to contribute to the present issue of OSID on the
quantum coding theorem. 

\begin{appendix}

\section{Unitary invariant Hermitian form}\label{app-unitary-invariance}
Let $H$ be a finite dimensional Hilbert space with an orthonormal
basis $\ket 1, \dots, \ket N$, and let $b:\cl B(H)\times \cl B(H) \to
{\mathbf C}$ be a unitary invariant Hermitian form. For $i,j \in \{1,\dots,
N \}$ let $E_{ij} := \ket{i} \bra{j}$. 
As a consequence of the unitary invariance one finds constants
$\alpha, \beta$ and $\gamma$ such that for 
$i,j,\in \{ 1, \dots, N \}$, $i\neq j$
\bea
b(E_{ij}, E_{ij}) &=& \alpha\:,\\
b(E_{ii}, E_{jj}) &=& \beta\:, \\
b(E_{ii}, E_{ii}) &=& \gamma\:,
\eea
and for all other combinations of indices $i,j,l,m \in \{ 1, \dots, N
\}$
\be
b(E_{ij}, E_{lm}) = 0\:.
\ee
This immediately leads to 
\be
b(V,W) = (\gamma- \alpha-\beta)\: b_1(V,W) \: + \: \beta \:\tr V\dag \tr W\: +
\: 
\alpha \:\tr\: V\dag W\:,
\ee
with 
\be
b_1(V,W) = \sum_{i=1}^N \sp{i|V\dag| i} \sp{i| W | i}\:.
\ee
Obviously, $b_1$ is not unitary invariant, from which we conclude
$\gamma-\alpha-\beta=0$ and thus
\be
b(V,W) = \beta\: \tr V\dag \tr W\: + \: 
\alpha \: \tr\: V\dag W\:,
\ee
which is what we wanted to prove.

\section{Average of $| \sp{\psi|\pi_C|\psi}|^2  $} \label{app-rmt}

We show that independent of the normalized vector $\ket \psi \in Q$
\ben\label{to_prove}
\cav{| \sp{\psi|\pi_C|\psi}|^2 }_{K} = \fr{1 + K\inv}{M^2 + M}
\een
(notations as in Sec.\ \ref{sec-average}).
By definition,
\be 
\cav{| \sp{\psi|\pi_C|\psi}|^2 }_{K} = \fr{1}{K^2}
\int d\mu(U) \: | \sp{\psi| U \:\Pi_0   U\dag |\psi}|^2\:,
\ee
where the integral extends over $\b U(Q)$ and $\Pi_0$ is the projection
on an arbitrarily chosen linear subspace $C_0 \subset Q$ of
dimension $K$. We extend $\ket \psi \equiv \ket{\psi_1}$ to an orthonormal basis 
$\ket{ \psi_1 }, \dots, \ket{ \psi_M }$ of $Q$, and chose
\be
C_0 := \span\{ \ket{ \psi_1 }, \dots, \ket{ \psi_K }\}\:.
\ee
Then
\be
\int d\mu(U) \: | \sp{\psi| U\: \Pi_0   U\dag |\psi}|^2
= \sum_{i,j=1}^K \int d\mu(U) \: 
|U_{1i}|^2 |U_{1j}|^2 \:,
\ee
where $U_{ij} = \sp{ \psi_i | U | \psi_j}$. Making use of the unitary
invariance of $\mu$, this becomes
\be
 K \int d\mu(U)\: |U_{11}|^4 
\: + \: 
(K^2-K) \int d\mu(U) \: |U_{11}|^2 |U_{12}|^2\:.
\ee
For the calculation of these integrals we refer to the work of Pereyra
and Mello \cite{PM83}, in which, amongst others, the joint probability
density for the elements  
$U_{11}, \dots, U_{1k}$ of a random unitary matrix $U\in U_K$ has been
determined to be  
\be 
p(U_{11}, \dots, U_{1k}) = c 
\left(1 - \sum_{a=1}^k |U_{1a}|^2 \right)^{n-k-1} 
\Theta(1 - \sum_{a=1}^k |U_{1a}|^2) \:, 
\ee
where $c$ is a normalization constant, and $\Theta(x)$ denotes the
standard unit step function. 
By a straightforward calculation, we obtain from this 
\bea
\int d\mu(U)\: |U_{11}|^4  &=& \fr{2}{M^2 + M}\:, \\
\int d\mu(U) \: |U_{11}|^2 |U_{12}|^2 &=& \fr{1}{M^2 + M}\:,
\eea
which immediately leads to Eq.\ \Ref{to_prove}.

\section{Typical Sequences}\label{app-typical}
The first property follows from 
\be 
1  =  
\sum_{\mbox{$\b a \in \aleph^n$}} p_{\b a} 
\: \ge \:
\sum_{\mbox{$\b a \in \aleph_{\eps,n}$ }} p_{\b a} 
\: \ge \:
|\aleph_{\eps,n}| 2^{-n(H(\cl P) +\eps)}\:.
\ee
To prove the second property we first realize that by definition
\bea
P_{\eps,n} 
&=&\Pr( \: \mbox{``$\b a \in \aleph^n$ is $\eps$-typical''} \:) 
\: = \: \Pr( \left|-\log_2(p_{\b a}) - n H(\cl P) \right| \:
\le\: n \eps) \\
&=& \Pr(\: | \sum_{l=1}^n \left(-\log_2 \cl P(a_l) - H(\cl P)\right) |\: 
\le\: n \eps\:)\:. 
\eea
The negative logarithms of the probabilities $\cl P(a_l)$ can be
understood as $n$ independent random variables $Y_l$
that assume values $-\log_2 \cl P(a)$ for all $a\in \aleph$ with probabilities
$\cl P(a)$. Their mean is the Shannon entropy $H(\cl P)$, 
\be
\mu = E(Y_1) = - \sum_{a\in \aleph} \cl P(a) \log_2 \cl P(a) = H(\cl P)\:.
\ee
This means that 
\be 
1 - P_{\eps,n}  
\: = \: \Pr( \: |\sum_{l=1}^n (Y_l - \mu)| \: \ge \: n \eps \:)
\ee
is the probability of a large deviation $\propto n$.
Since the variance $\sigma$ and all higher
moments of $Y_1-\mu$ are finite we can employ a result from the theory 
of large deviations \cite{GS92}, according to which 
\be
\Pr(\: |\sum_{l=1}^n (Y_l - \mu)| \: \ge \: n \eps \:)
\: \le  \: 2 e^{ -n \psi(\eps)}\:, 
\ee
where $\psi(\eps)$ is a positive number that is
approximately $\eps^2 / 2 \sigma^2$.

\section{Properties of $\cl   N_{\eps,n}$
 and  $ \tilde \cl   N_{\eps,n}$ }\label{app-properties}  
We will show the following relations
(definitions and notations as in Sec.~\ref{subsub-reduction}):
\def \neps{ {\cl N_{\eps,n}} }
\def \tneps{ {\tilde\cl N_{\eps,n }}}
\bean
|\neps| & \le & 2^{n\left(S_e(\pi, \cl N) + \eps\right)} \label{1}\\
\tr\:\neps(\pi_n) & \ge & 1 - 2 e^{-n \psi_1(\eps)} \label{2}\\
|\tneps| & \le & 2^{n\left(S_e(\pi, \cl N) + \eps\right)} \label{3}\\
\tr\:\tneps(\pi_n) & \ge & 1 - 4 e^{-n \psi_3(\eps)} \label{4}\\
\fn{ \tneps(\pi_n)}^2 & \le & 2^{-n\left( S(\cl N(\pi)) - 3 \eps\right)}\:,\label{5}
\eean
where $\psi_1(\eps)$ and $\psi_3(\eps)$ are positive numbers
independent of $n$.

The first relation follows from 
$
|\neps| = |\aleph_{\eps,n}| \le 2^{n \left( H(\cl P) + \eps \right)}
$
and $H(\cl P) = S_e(\pi, \cl N)$. 
To prove relation \Ref{2} we note that for a Kraus operator $\b A =
A_{j_1} \otimes \dots \otimes 
A_{j_n}$
\be
\fr{1}{|Q|^n} \tr\: \b A\dag \b A 
=
\fr{1}{|Q|} \tr\: A_{j_1}\dag A_{j_1} \dots 
\fr{1}{|Q|} \tr\: A_{j_n}\dag A_{j_n}
=
\cl P(A_{j_1}) \dots \cl P(A_{j_n}) 
\equiv p_{\b A}\:.
\ee
Making use of property (ii) of typical sequences this shows
\be
\tr\: \cl N_{\eps,n}(\pi_n) = \fr{1}{|Q|^n} \sum_{\b A\in
  \aleph_{\eps,n}} \tr\: \b A\dag \b A = \sum_{\b A\in 
  \aleph_{\eps,n}} p_{\b A} \ge 1 - 2 e^{n \psi_1(\eps)}\:,
\ee
where $\psi_1(\eps)$ is a positive number independent of $n$. 
Relation \Ref{3} is evident by relation \Ref{1} and 
\be
\tneps(\rho) = \Pi_{\eps,n} \: \neps(\rho)\: \Pi_{\eps,n} 
= 
\sum_{\b A \in
  \aleph_{\eps,n}} (\Pi_{\eps,n} \b A) \rho  
(\Pi_{\eps,n} \b A)\dag\:.
\ee

In order to show \Ref{4} it is convenient to
introduce the complementary operation $\cl M_{\eps,n}$ of $\cl
N_{\eps,n}$ by 
\be 
\cl N\tenpow n = \cl N_{\eps,n} + \cl M_{\eps,n}\:,
\ee 
i.e. $\cl M_{\eps,n}$ consists of the 
$\eps$-``untypical'' Kraus operators of $\cl N \tenpow n$, 
\be
\cl M_{\eps,n}(\rho) = \sum_{\b A \in \aleph \setminus
  \aleph_{\eps,n}} \b A \: \rho \: \b A\dag\:.
\ee
Then, 
\bean
\tr\: \tilde{\cl N}_{\eps,n} (\pi_n ) 
&=&
       \tr\: \Pi_{\eps,n}( \cl N \tenpow n (\pi_n)-\cl M_{\eps,n}
       (\pi_n)) \nn\\ 
& \ge&  \tr  \Pi_{\eps,n} \cl N \tenpow n (\pi_n) 
     - \tr\: \cl M_{\eps,n} (\pi_n)\:. \label{tr-tildeN-inequality}
\eean
The inequality results from the fact that for two positive operators 
$A,B$ always $\tr AB \ge 0 $, and therefore (indices suppressed)
\be
\tr\: \cl M(\rho)= \tr\: \Pi \cl M(\rho) + \tr\: (\b 1 - \Pi) \cl M(\rho) \ge
\tr\: \Pi \cl M(\rho)\:. 
\ee
Taking into account that $\Pi_{\eps,n}$ projects on the typical subspace
$T_{\eps,n}$ of $\cl N(\pi)$ and using property (ii') of typical
subspaces, the first term in Eq.\ \Ref{tr-tildeN-inequality} can be
bounded from below as 
\be
\tr\: \Pi_{\eps,n} \cl N \tenpow n (\pi_n) 
 = 
\tr\: \Pi_{\eps,n} \cl N \tenpow n(\pi \tenpow n)  
 = 
\tr\: \Pi_{\eps,n} (\cl N(\pi) )\tenpow n 
\: \ge \:
1 - 2 e^{-n \psi_2(\eps)}\:. 
\ee
The second term in Eq.\ \Ref{tr-tildeN-inequality} obeys
\be
\tr\: \cl M_{\eps,n}( \pi_n )
\: = \:
\tr \: \cl N \tenpow n ( \pi_n) \: - \: \tr \: \cl N_{\eps,n}(
\pi_n) \: \le \: 2 e^{ -n \psi_1(\eps)}\:, 
\ee
by relation \Ref{2}. We thus find
\be
\tr\: \tilde{\cl N}_{\eps,n} (\pi_n )  
\: \ge \: 
1 - 2 (e^{-n \psi_2(\eps)} + e^{ -n \psi_1(\eps)}) 
\: \ge \:
1 - 4\: e^{-n \psi_3(\eps)}
\:,
\ee
when $\psi_3(\eps) := \min\{ \psi_1(\eps), \psi_2(\eps) \}$.

Finally, we address the Frobenius norm of $\tilde{\cl N}(\pi_n)$. 
For positive operators $A,B$
\be
\fn{ A + B}^2  
\: =   \: \fn{A}^2 + \fn{B}^2 + 2 \tr\: A B
\: \ge \: \fn{A}^2 + \fn{B}^2\:. 
\ee
This can be used to derive 
\be
\fn{ \cl T_{\eps,n} \circ \cl N \tenpow n (\pi_n) }^2 
\: = \:
\fn{ \cl T_{\eps,n} \circ ( \cl N_{\eps,n} + \cl
  M_{\eps,n})(\pi_n)}^2 
\: \ge \:
\fn{ \cl T_{\eps,n} \circ \cl N_{\eps,n}(\pi_n)}^2\:. 
\ee
Thus
\bea
\fn{ \tilde{\cl N}_{\eps,n} (\pi_n) }^2 
& = &
\fn{ \cl T_{\eps,n} \circ \cl N_{\eps,n}(\pi_n)}^2 \\
& \le &
\fn{  \cl T_{\eps,n} \circ \cl N \tenpow n (\pi_n) }^2 \\
& = &
\fn{ \Pi_{\eps,n} \: ( \cl N(\pi) )\tenpow n \: \Pi_{\eps,n} }^2 \\
& = &
\sum_{\scrbox{ $ \ket{\b v}$ $\eps$-typical eigenvector} }
\left( p_{\b v} \right)^2 \\
&\le & 2^{-n ( S( \cl N(\pi)) - 3 \eps)}\:, 
\eea
where we used $\dim T_{\eps,n} \le 2^{n( S( \cl N(\pi)) + \eps)}$
(property (i')) and $ p_{\b v} \le 2^{-n( S( \cl N(\pi)) -
\eps)}$ to derive the last inequality. 

\end{appendix}

\end{document}